\documentclass[preprint]{aastex}

\shorttitle{GRB-SNe as standardizable candles}
\shortauthors{Z. Cano}

\begin{document}

\title{Gamma-ray burst supernovae as standardizable candles}

\author{Z. Cano}
\affil{Centre for Astrophysics and Cosmology, Science Institute, University of Iceland, Reykjavik, Iceland.}
\email{zewcano@gmail.com}

\begin{abstract}
A long-duration gamma-ray burst (GRB) marks the violent end of a massive star.  GRBs are rare in the universe, and their progenitor stars are thought to possess unique physical properties such as low metal content and rapid rotation, while the supernovae (SNe) that are associated with GRBs are expected to be highly aspherical.  To date, it has been unclear whether GRB-SNe could be used as standardizable candles, with contrasting conclusions found by different teams.  In this paper I present evidence that GRB-SNe have the potential to be used as standardizable candles, and show that a statistically significant relation exists between the brightness and width of their decomposed light curves relative to a template supernova.  Every single nearby spectroscopically identified GRB-SN, for which the rest-frame and host contributions have been accurately determined, follows this relation.  Additionally, it is shown that not only GRB-SNe, but perhaps all supernovae whose explosion is powered by a central engine, may eventually be used as a standardizable candle.  Finally, I suggest that the use of GRB-SNe as standardizable candles likely arises from from a combination of the viewing angle and similar explosion geometry in each event, the latter which is influenced by the explosion mechanism of GRB-SNe.
\end{abstract}

\keywords{TBC}

\section{Introduction}
The discovery of the energetic, stripped-envelope core-collapse SN~1998bw, which was temporally and spatially coincident with GRB~980425 (Galama et al. 1998; Patat et al. 2001), provided the first empirical evidence that GRBs arise from massive stars.  Since 1998 there have been several more spectroscopic (e.g. Pian et al. 2006; Sparre et al. 2011; Bufano et al. 2012; Xu et al. 2013; Schulze et al. 2014) and photometric observations (e.g. Stanek et al. 2005; Bersier et al. 2006; Cano et al. 2011B) of supernovae (SNe) accompanying GRBs (GRB-SNe).  GRB-SNe are very energetic, possessing, on average, ten times more kinetic energy than SNe Ibc not associated with GRBs (Iwamoto et al. 1998).  Every single GRB-SNe that has been spectroscopically classified is of type Ic, implying that their outer layers of hydrogen and helium have been lost prior to explosion.  Moreover, the blueshifted line velocities in their spectra are much larger than those measured for non-GRB SNe Ibc, approaching velocities more than 10\% the speed of light, and which display broad-lined (BL) absorption lines.  Such rapid line velocities and broad-lined profiles have been seen in the spectra of SNe Ic-BL not associated with GRBs, which are also more energetic and have larger ejecta velocities than SNe Ibc, though not as extreme as GRB-SNe (Cano 2013).

The exact nature of the central engine in GRB-SNe is uncertain, though theory suggests it may being either an accreting black hole, or a rapidly rotating neutron star that possesses a very strong magnetic field.  In these models (e.g. Woosley 1993; MacFadyen et al. 1999; MacFadyen et al. 2001), a stellar black hole is created during the core-collapse of a rapidly rotating massive star (e.g. Yoon \& Langer 2005; Woosley \& Heger 2006).  The angular momentum present in the core at the time of collapse leads to the formation of a torus of material that accretes onto the BH, ultimately producing an ultra-relativistic, bi-polar jet along the rotation axis.  Once the jets pierce through the star, they interact with themselves producing the initial burst of $\gamma$-rays, and later they collide with the surrounding medium, producing the longer-lasting X-ray/optical/radio afterglow (AG).  It has been suggested by Bromberg et al. (2011) that low-luminosity GRBs such as GRB~060218 and GRB~100316D arise from jets that just barely breakout from the stellar surface, and therefore emit less $\gamma$-ray radiation.  

Adding diversity to this scenario is the existence of type Ic-BL SN~2009bb (Soderberg et al. 2010; Pignata et al. 2011).  To date, SN~2009bb is one of only two SN Ic-BL (the other being SN~2012ap; Margutti et al. 2014; Milisavljevic et al. 2014) not associated with a GRB for which there is compelling observational evidence that they both possessed a central engine that contributed to the explosion of their stellar progenitors.  Luminous radio detections of SN~2009bb were attributed by Soderberg et al. (2010) as arising from a substantially relativistic outflow powered by a central engine.  The kinetic energy contained within the relativistic outflow of SN~2009bb was demonstrated to be similar to that of low-luminosity GRBs 060218 and 100316D (Margutti et al. 2013; Margutti et al. 2014), which indirectly suggests that SN~2009bb may have a similar degree of collimation/asymmetry as low-luminosity GRB-SNe.  In the above theoretical scenario, SN~2009bb arises from a jet that failed to breakout of the star, possibly because of a less energetic central engine.   

Another open question is whether GRBs and/or GRB-SNe can be used as standardizable candles.    While a few GRB-SNe have occurred in the local universe, with the relative brightness of the event allowing a direct analysis of their observational and physical properties to be performed, isolating the SN contribution from the optical and near-infrared (NIR) observations of fainter events is more complicated.  Precisely ascertaining the observational properties of the accompanying SNe requires a careful decomposition of the photometric observations.  The degree of accuracy obtained in decomposition methods depends strongly on precisely knowing the total amount of extinction along the line of sight (both foreground and rest-frame), where determining the latter requires assumptions that complicate the analysis.  Furthermore, the contribution from the host galaxy needs to be quantified and removed, either by image-subtraction techniques, or by mathematically subtracting the host flux from the observations.  Finally, in order to isolate and model the flux from the SN, the temporal behaviour of the optical AG, which arises from the initial GRB event, needs to be precisely determined.  If the AG is improperly modelled, the AG contribution can be under- or over-estimated, which affects the resultant SN properties.  Therefore, in order for the SN to be modelled accurately, each one of these hurdles needs to be carefully overcome.

Several studies (Zeh et al. 2004; Ferrero et al. 2006; Cano et al. 2011A; Th{\"o}ne et al. 2011; Cano 2013) have attempted to determine the luminosity ($k$) and stretch ($s$) factor of GRB-SNe relative to a template SN, which historically has been SN~1998bw.  These parameters are analogous to the absolute peak SN magnitude and $\Delta m_{15}$ (the amount the light curve fades from peak light to fifteen days later) used in the luminosity--decline relation for SNe Ia (Phillips 1993).  Initial studies such as those by Ferrero et al. (2006) and Cano et al. (2011A) noted that no obvious correlation exist in the observer-frame $R$-band $k$- and $s$-factors of several GRB-SNe.  However, as the GRB-SNe in their respective samples occurred over a broad range of redshifts (0.0085$\le z \le1.006$), the $R$-band light curves (LCs) in their respective analyses were all probing different rest-frame wavelengths, and thus different regions of their spectral energy distributions (SEDs).  Encouragingly, a recent analysis by Schulze et al. (2014) looked at the \textit{approximate} rest-frame $V$-band stretch and luminosity factors of several GRB-SNe, finding that a rough correlation was present.  An additional analysis by Lyman et al. (2014) also hinted at a correlation between the peak absolute magnitude and $\Delta m_{15}$ of several nearby GRB-SNe and SN~2009bb.  Excitingly, a very recent paper by Li \& Hjorth (2014) have presented a luminosity--decline correlation between the K-corrected\footnote{Not to be confused with the $k$ parameter used here to denote the brightness of a given GRB-SNe in a given filter, but it is rather the correction that arises from observing events occurring at vast cosmological distances, where the observer-frame light arises from bluer rest-frame light that has been redshifted as it traverses through an expanding universe.}, $V$-band, peak absolute magnitude and decay rates from peak of eight GRB-SNe.  Reassuringly, their results fully support those presented in this paper.

In this present work I performed a detailed analysis to precisely account for the varying redshifts of GRB-SNe in order to determine the \textit{exact} rest-frame ($UBVRI$) stretch and luminosity factors of several GRB-SNe.  The latest values of the observer-frame (foreground) extinction due to the Milky Way (MW) from Schlafly \& Finkbeiner (2011) have been used, as well as values for the rest-frame extinction derived by several works (see the Appendix for a detailed discussion of each GRB-SN).  Importantly, this analysis is an improvement over previous works (e.g. Cano et al. 2011A; Cano 2013) by calculating the \textit{redshift-corrected} rest-frame extinction using the value of $A_{V,\rm rest}$, and the MW/SMC/LMC templates of Pei (1992).  This is more precise than considering a total line-of-sight extinction for a given filter, for at larger redshifts, the extinction measured in a certain observer-frame filter will arise from bluer light in the rest-frame that will be correspondingly more extinguished.  A $\Lambda$CDM cosmology constrained by Planck (Planck Collaboration et al. 2013) has been employed here, where $H_{0} = 67.3$ km s$^{-1}$ Mpc$^{-1}$, $\Omega_{M} = 0.315$, $\Omega_{\Lambda} = 0.685$.

\section{Procedure}

\subsection{Decomposing the optical light curves}

\begin{figure}
 \centering
 \includegraphics[bb=0 0 448 281]{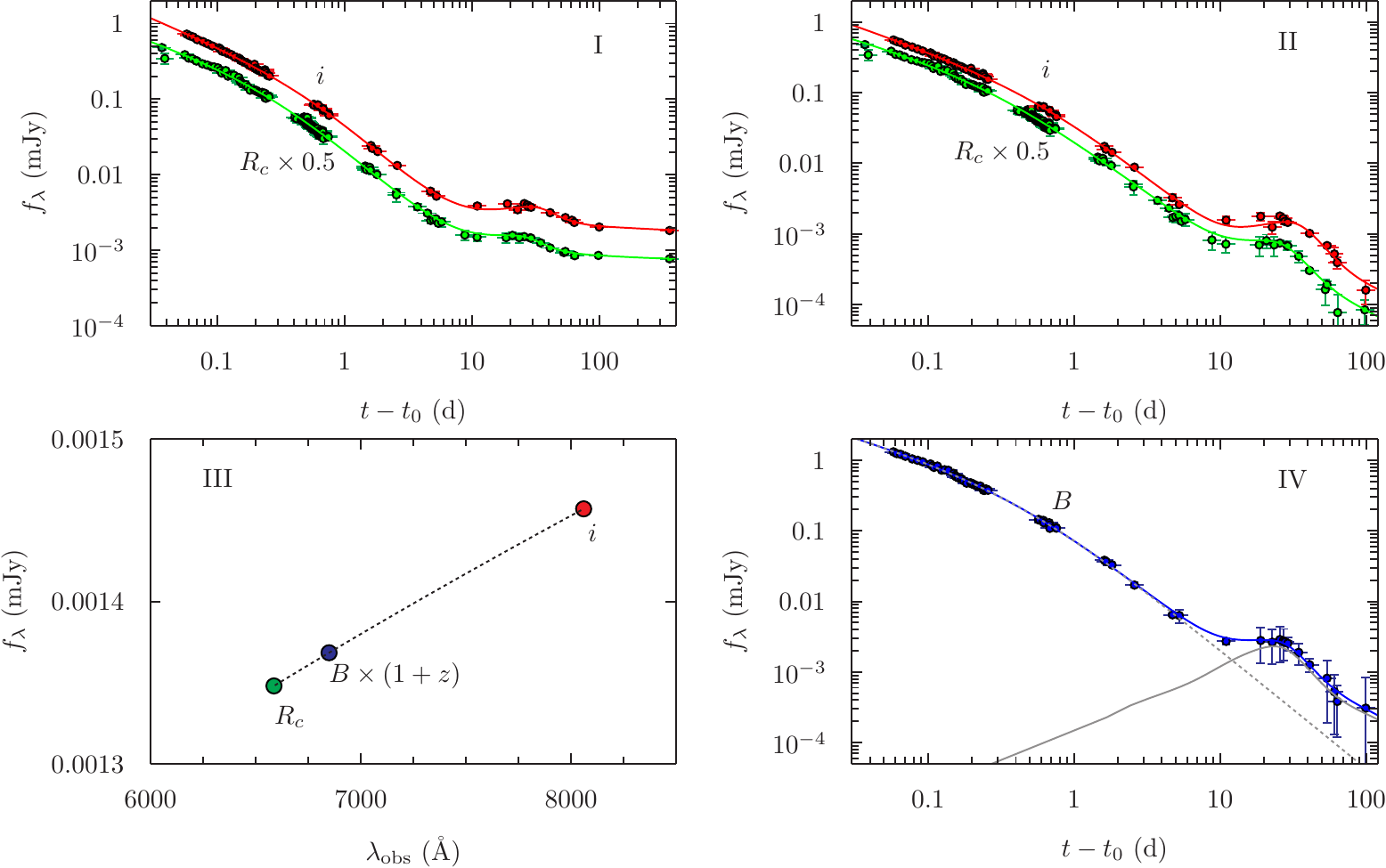}
 % 090618.pdf: 448x281 pixel, 72dpi, 15.80x9.91 cm, bb=0 0 448 281
   \caption[]{\small  Decomposition of the optical LC of GRB 090618.  \textit{Upper Left} (I): Original LCs of GRB~090618 in $R$-band (the flux has been multiplied by a factor of 0.5 for clarity; in green) and $i$-band (red), which have been corrected for observer-frame (foreground) extinction.  The solid lines in both colours are the best-fits to the AG+SN+host.    \textit{Upper Right} (II): same as panel I, but the host flux has been mathematically subtracted.  The SN ``bump'' is more clearly pronounced.  \textit{Lower Left} (III): an example SED, where the observer-frame $R$- and $i$-bands are shown in green and red, respectively.  The SED has been fit with a log-linear spline (dashed black line), which has been interpolated to $\lambda_{B,\rm rest} \times (1+z)$.  \textit{Lower Right} (IV):  Rest-frame $B$-band redshifted to observer-frame $z=0.54$ (blue), and corrected for rest-frame extinction.  The grey, dashed line is the model of the AG ($\alpha_{1}=0.72\pm0.19$, $\alpha_{2}=2.04\pm0.12$, and time the LC breaks from $\alpha_{1}$ to $\alpha_{2}$ is $1.26\pm0.49$ d).  The solid grey line is our template supernova (SN 1998bw).  The solid line is the sum of the AG and SN model, where the luminosity and stretch factor was found to be $k=1.11\pm0.22$ and $s=0.98\pm0.20$ respectively.  The errors in $k$ and $s$ result from the propagation of the various individual sources of uncertainty (see main text), which are estimated to be of order 20\%.}
 \label{fig:090618}
\end{figure}

The method used involves decomposing the optical LCs to isolate flux coming from just the accompanying SN.  Previous papers (e.g. Zeh et al. 2004; Ferrero et al. 2006; Cano et al. 2011A; Th{\"o}ne et al. 2011; Cano 2013) have used similar methods to determine the brightness ($k$) and width/stretch factor ($s$) in a particular filter relative to a template SN, which historically has been SN~1998bw due to its well-sampled LC and well-determined temporal evolution.  A step-by-step visual diagram of the decomposition method is displayed in Fig. \ref{fig:090618}. 

For each GRB-SN in the sample, I collected all of the available data published in the literature (Table \ref{table:refs}).  While the amount of SN-bumps in optical and NIR LCs is more than 20 (Cano 2013), the analysis has been limited to those events where: (1) observations were in at least two filters in order to interpolate between filters to get to the precise to rest-frame wavelength), (2) observations of the host had been obtained, (3) knowledge of the rest-frame extinction had been determined, and (4) the LC is well sampled enough that the AG and SN temporal behaviour can be accurately and, importantly, unambiguously modelled.  Using this criteria, eight GRB-SNe were analysed, not including the template SN~1998bw.  Once the data were gathered, the magnitudes were corrected for foreground extinction and then converted into monochromatic fluxes, using the effective wavelengths and flux zeropoints in Fukugita et al. (1995).

A vital part of the decomposition is quantifying the total extinction along the line of sight.  Observer-frame (foreground) extinction due to various sight-lines through the Milky Way was corrected for using the dust maps of Schlafly \& Finkbeiner (2011).  The rest-frame extinction is more complex to determine.  One way is to obtain a spectrum of the host galaxy after the GRB and SN have faded away, and use either absorption or emission line diagnostics to estimate the host-galaxy extinction.  The draw-back of this method is that for GRBs at higher redshift that cannot be resolved in their host galaxy, one has to make the assumption that the integrated dust extinction of the entire galaxy is a suitable proxy for the dust-extinction local to the GRB event itself, which for some GRB events has proved to be a poor proxy (e.g. Kr{\"u}hler et al. 2011).  Another method is to model early observations of the AG at optical, IR and X-ray frequencies, and fit the data with extinction curves based on the SMC, LMC and MW (e.g. Kann et al. 2006; Guidorzi et al. 2009; Kann et al. 2010; Cano et al. 2011B; Olivares et al. 2012).  The rest-frame extinction for the majority of the GRB-SNe in the sample has been determined using the latter method.  Note that the rest-frame extinction is only corrected for after (1) the observer-frame host contribution has been removed, and (2) the interpolation described below was completed.

For each GRB-SN, the light is attributed as coming from three different sources: (1) the AG, which is associated with the GRB event, (2) the SN, and (3) the constant source of flux coming from the host galaxy.  The host contribution was determined and removed either by image subtraction (e.g. GRB 011121; Bloom et al. 2002), or by observing the host galaxy at late times after the AG and SN have faded away, and then mathematically subtracting the host flux from the earlier epochs (e.g. GRB 130831A; Cano et al. 2014).  Note that the host galaxy contribution needs to be removed at this stage, because once the SED interpolation is performed, we do not have knowledge of the brightness of the host galaxy at the redshift-corrected, observer-frame wavelength to later remove.  After removing the host contribution, the remaining data is attributed to coming from just the AG and SN.  

At this point, observer-frame SEDs at each epoch of contemporaneous optical data for each GRB-SN are created.  For example, for GRB 090618 (redshift $z=0.54$), I have used observer-frame $R_{c}$ and $i$ data from Cano et al. (2011B), which using the effective wavelengths of each filter from Fukugita et al. (1995) corresponds to observer-frame wavelengths of 6588 and 7706 \AA $ $ respectively.  Rest-frame $B$-band is 4448 \AA, which corresponds to observer-frame 4448($1+z$) $=$ 6850 \AA.  Therefore at each epoch of host-subtracted data,  the SED has been interpolated to this wavelength to extract the flux at redshifted rest-frame $B$-band.  

Once the redshift-corrected LC is created, which has now been corrected for rest-frame extinction, it needs to be modelled to determine the temporal behaviour of the AG, as well as the properties of the accompanying SN.  Standard GRB theory suggests that the light powering the AG is synchrotron in origin, and therefore follows a power-law behaviour in both time and frequency ($f_{\nu} \propto (t - t_{0})^{-\alpha}\nu^{-\beta}$, where the respective decay and energy spectral indices are $\alpha$ and $\beta$, while $t_{0}$ is the trigger time of the GRB).  It is therefore very beneficial to have a wide time coverage of the entire event in order to accurately determine the rate at which the AG decays, especially at late times when the SN is the dominant source of flux.

Finally, in order to measure the luminosity and stretch factors of each accompanying SN, I used a redshifted (K-corrected) template of SN 1998bw, originally presented in Cano (2013).  The templates are created using a \textsc{c}-program that calculates the rest-frame filter LCs for each SN and interpolates in frequency and time over the original LCs of SN~1998bw (Galama et al. 1998; Patat et al. 2001), and then calculates the luminosity distance using a $\Lambda$CDM cosmology to rescale the flux density.  The program generates a synthetic LC at the desired observer-frame wavelength.

The synthetic LC is then fitted with a log-linear spline using \textsc{pyxplot}\footnote{http://pyxplot.org.uk}.  If we denote the log-linear spline as function $\Lambda(x)$, then to determine the stretch and luminosity factors of each SN relative to the template we use a new function of the form $\Sigma = k \times \Lambda(x/s)$.  This function is added to the AG function, and the data is fitted to simultaneously determine the decay rate of the AG at all times, as well as $k$ and $s$, all of which are free parameters.  It is seen that the LC of GRB~090618 has a ``break'' in it, where a steepening of the AG occurs at some break time ($T_{\rm B}$).  Thus, two additional free parameters are included ($T_{\rm B}$ and the decay constant after the break).  A summary of the best-fit parameters are presented in Table \ref{table:params}.

Throughout this analysis there are several sources of errors: host-galaxy subtraction, afterglow subtraction, uncertainties in the observer-frame and rest-frame extinction and the SED construction and interpolation.  Each one of these procedures/quantities introduces an error into the final calculation of $k$ and $s$.  The errors of the first four are photometric, and the errors have been added in quadrature.  A conservative estimation of the total error in $k$ and $s$, which includes an estimate of the error introduced by the SED interpolatiom, is of order 20\%, and is the fractional value adopted in Table \ref{table:params}.

A visual example of the error propagation can be seen in Fig. \ref{fig:090618}, where the final errorbars of the rest-frame $B$-band LC are larger than those of the observer-frame $R$- and $i$-band LCs as they also include the uncertainty associated with the rest-frame extinction.

\section{Results}

For every GRB-SNe in this sample, I interpolated the optical/NIR spectral energy distributions (SEDs) to exact rest-frame frequencies, and then performed the LC decomposition method to determine $k$ and $s$ at every possible rest-frame filter.  The resultant value of $k$ and $s$ in each rest-frame filter provides an independent measure of the shape of the SN relative to the template SN.  Note that this approach can be considered to be model-dependent as it relies on using a template SN, rather than focusing on the SN properties in and of themselves.  As such, I have assumed that each GRB-SN evolves in a similar fashion to the template SN, corrected only for the unique intrinsic brightness and temporal evolution of each GRB-SN, but with similar color evolution.  It has been shown by e.g. Lyman et al. (2014) that GRB-SNe, and all SNe for that matter, will possess a unique colour evolution.  However, it should be noted that a GRB-SN with $k$ and $s$ values that are not identical do account for their unique colour evolution relative to the template SN.  For example, a fiducial event with $s_{R}=0.8$ and $s_{I}=0.9$ implies that the SN gets redder over time more rapidly than an event with $s_{R}=s_{I}=0.9$ (in the former event, $R-I$ will increase more rapidly after peak light as the SN fades more rapidly in $R$ relative to $I$).  As such, the method used here does provide a way to account for the unique colour evolution of each GRB-SN.

With this caveat in mind, the analysis of Li \& Hjorth (2014) can be considered to be less model dependent in this respect, though it should be considered that their analysis still relies on using the SED of SN~1998bw as a template when calculating the K-correction of those events in their sample where their constructed SED is incomplete.  

\begin{figure}
 \centering
 \includegraphics[bb=0 0 576 432,scale=0.8]{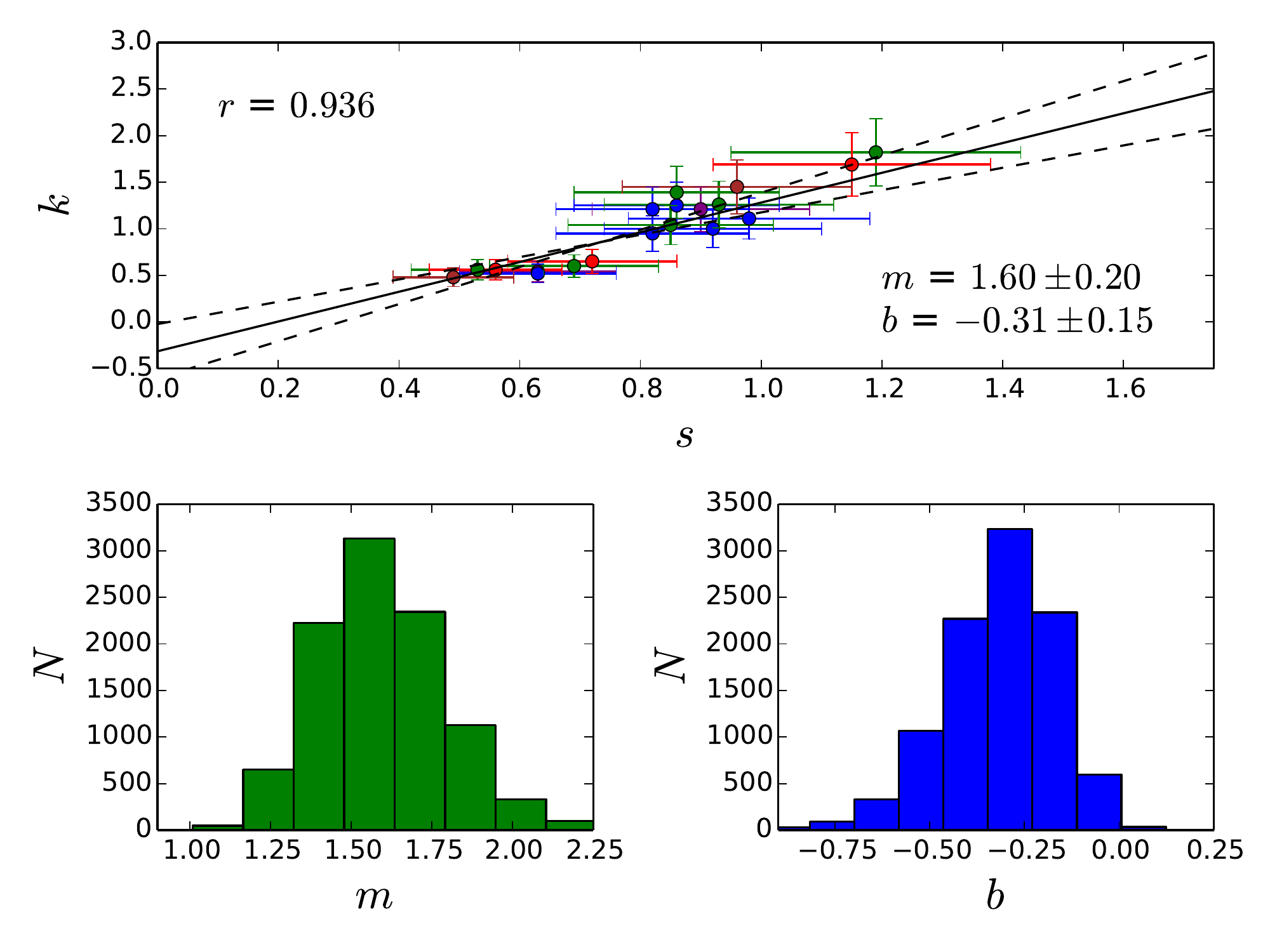}
 % Candles_LLS.pdf: 576x432 pixel, 72dpi, 20.32x15.24 cm, bb=0 0 576 432
 \caption{The luminosity $k$ and stretch $s$ factors of the GRB-SNe (not including the template SN~1998bw). {\it Top Panel:} Combined data in rest-frame filters $U$ (purple), $B$ (blue), $V$ (green), $R$ (red) and $I$ (brown) are shown.  The data were analysed with a linear-least squares method coupled with a Monte Carlo simulation to determine the slope and y-intercept, and their respective errors, of the fitted line.  The best-fitting parameters are $m=1.60\pm0.20$ and $b=-0.31\pm0.15$, where the quoted errors are 1-$\sigma$, and are the standard deviations of each variable (see below).  Also plotted are the 2-$\sigma$ uncertainty limits of the fit (dashed lines).  The Pearson's correlation coefficient of the combined data is $r=0.936$, and the two-point probability of a chance correlation is $p=4.2\times10^{-9}$, which indicates the potential use of GRB-SNe as standardizable candles.  {\it Bottom Panels:} the distribution of the fitted values of $m$ (green) and $b$ (blue).}
\label{fig:GRB-SNe}
\end{figure}

Next, using \textsc{python} programs that employ \textsc{numpy} and \textsc{scipy} algorithms, I performed a non-linear least squares analysis on the combined dataset to find the best-fitting values of the slope ($m$) and y-intercept ($b$) to a straight line that was fitted to the data.  The fit was performed with the \textsc{optimization.curve$\_$fit} routine from \textsc{scipy}, which uses the Levenberg-Marquardt algorithm to fit the line to the data. The analysis was designed to (necessarily) consider the error in both $k$ and $s$.  To obtain an estimate of the error in $m$ and $b$, I performed a Monte Carlo analysis, where a new dataset is created from the original by choosing a new value from a Gaussian that is centered on each original datapoint, and whose standard deviation in each direction is equal to the error in that variable.  The simulation was performed 10,000 times, each time fitting the straight line to the new dataset.  The estimated errors of $m$ and $b$ arise from their resultant distributions, where the quoted 1-$\sigma$ errors are the standard deviation of each distribution.  The best-fitting values are $m=1.60\pm0.20$ and $b=-0.31\pm0.15$.  Interestingly, this result implies that GRB-SNe do not follow a 1:1 ratio of $k$:$s$, rather $k$ increases more rapidly with larger $s$.  The results are displayed in Figure \ref{fig:GRB-SNe}.

It is confirmed that a statistically significant correlation is present in the data.  The Pearson's correlation coefficient of the sample ($N=9$) is $r=0.936$, while the two-point probability of a chance correlation is $p=4.2\times10^{-9}$.  It is worth noting that every single nearby, spectroscopically identified GRB-SN obeys this relation, including the ``Big Five'' GRB-SNe (GRBs 980425, 030329, 031203, 060218 and 100316D) and GRB~120422A.  The notable absentee is recent GRB 130427A, which is associated with SN~2013cq (Xu et al. 2013).  The exclusion of this event is dictated by its complex AG behaviour and uncertain host galaxy contribution (Perley et al. 2014), which, like GRB 030329, severely hinders our ability to precisely decompose the optical and NIR observations, and will most likely require more advanced decomposition methods such as those used by Deng et al. (2005) on the spectra obtained of GRB 030329.

\begin{figure}
 \centering
 \includegraphics[bb=0 0 576 432,scale=0.8]{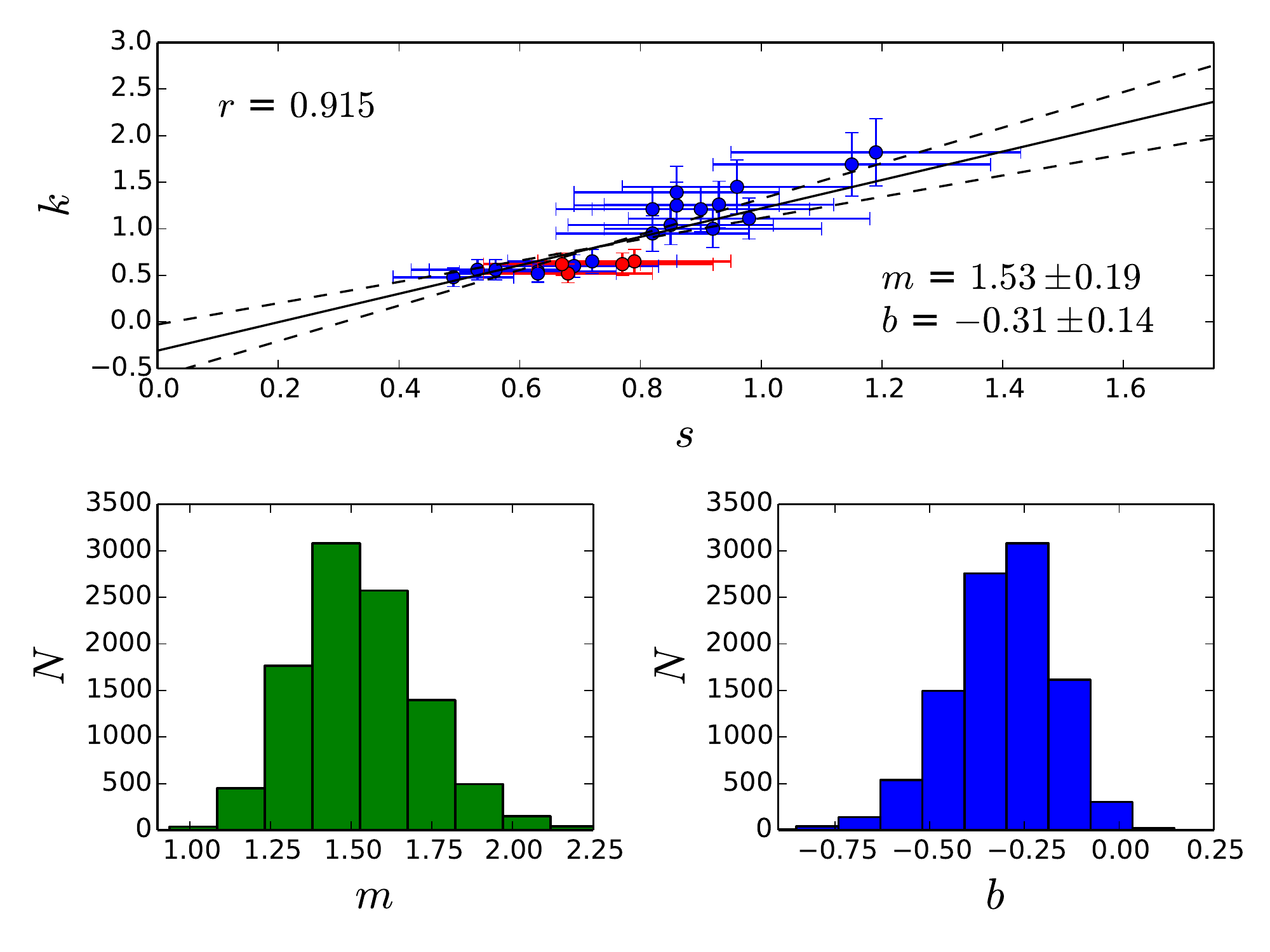}
 % Candles_IcBL_GRB.pdf: 576x432 pixel, 72dpi, 20.32x15.24 cm, bb=0 0 576 432
 \caption[]{Same analysis as Figure~1, but including the engine-driven Ic-BL SN~2009bb, which was not associated with a GRB.  When SN~2009bb (red) is included with the GRB-SN sample (blue), a Pearson's correlation coefficient of $r=0.915$ is found, a probability of a chance correlation of $p=9.7\times10^{-10}$, and best-fitting parameters of $m=1.53\pm0.19$ and $b=-0.31\pm0.14$.}
\label{fig:SNe_GRB_IcBL}
\end{figure}

Motivated by the observation that SN~2009bb is also a relativistic, type Ic-BL SN, I included its $BVRI$ stretch and luminosity factors from Cano (2013) with the GRB-SNe sample.  I have adopted an uncertainty of 20\% in the values of $k$ and $s$ for SN~2009bb in this analysis, which are larger than those originally presented in Cano (2013).  I then refit the data using the non-linear least squares and Monte Carlo analysis (Figure \ref{fig:SNe_GRB_IcBL}).  The Pearson's correlation coefficient for this new sample is $r=0.915$, with a chance probability of $p=9.7\times10^{-10}$.  Moreover, the values of $m=1.53\pm0.19$ and $b=-0.31\pm0.14$ are in good agreement with those determined from just the GRB-SN sample.  This result suggests that perhaps all engine-driven SNe are potential standardizable candles, though this conclusion is understandably tentative as it is based only on a single non-GRB-SN event.

\begin{figure}
 \centering
 \includegraphics[bb=0 0 576 432,scale=0.8]{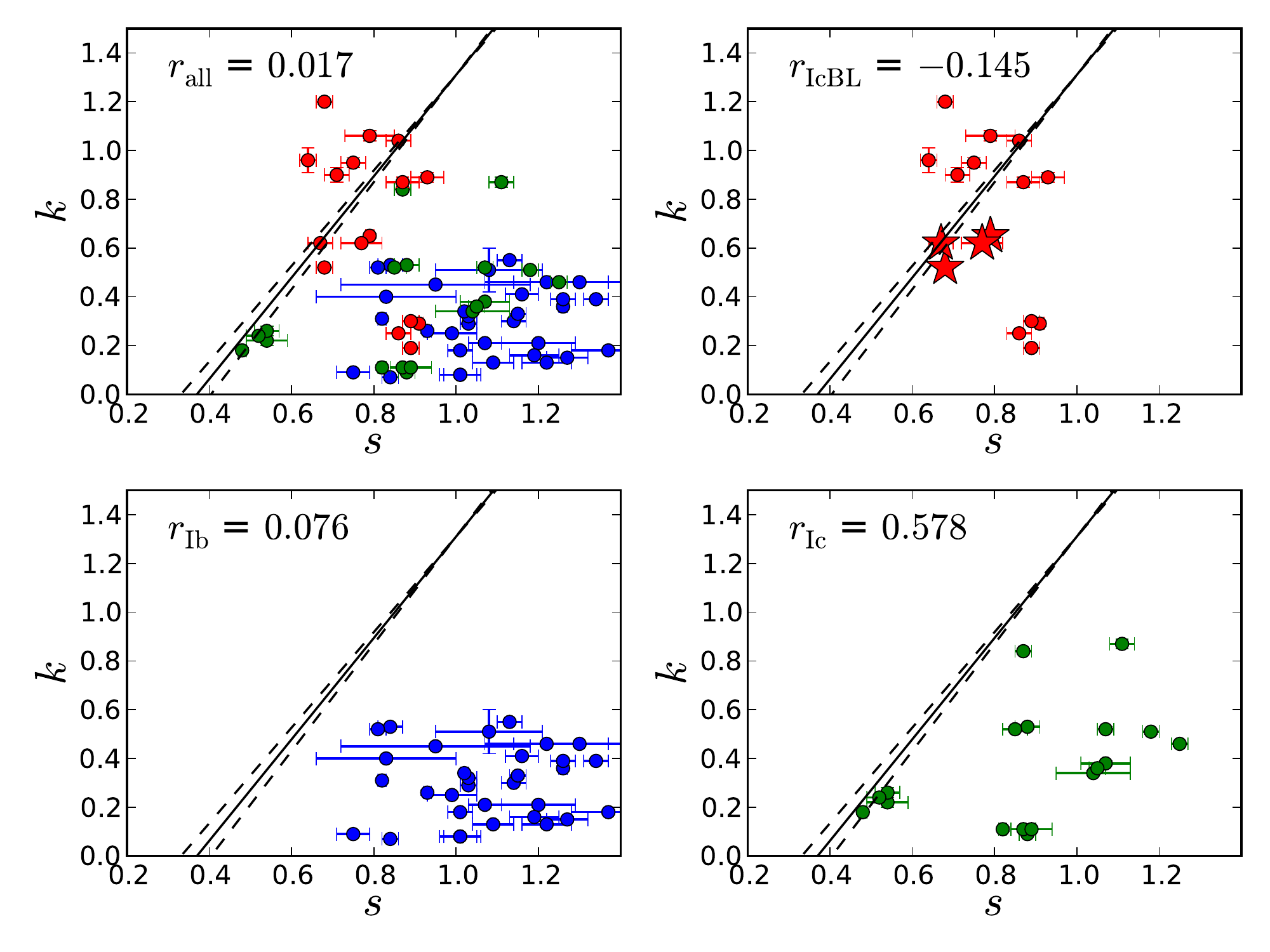}
 % Candles_all.pdf: 576x432 pixel, 72dpi, 20.32x15.24 cm, bb=0 0 576 432
 \caption[]{The luminosity $k$ and stretch $s$ factors of a sample of stripped-envelope supernovae (aka SNe Ibc) not associated with GRBs from \cite{Cano2013}, for which the total line-of-sight extinction is known and which are host corrected.  {\it Top Left:} The combined data of SNe Ib (blue), SNe Ic (green) and SNe Ic-BL (red).  Also plotted is the fit (solid line) derived for our GRB-SN sample along with the 2-$\sigma$ uncertainty limits (dashed lines).  The position of type Ic-BL SN 2009bb is shown in the upper right subplot as filled stars.  The Pearson's correlation of the combined SNe Ibc sample is $r_{\rm all}=0.017$, indicating they are not standardizable candles. {\it Remaining panels:} Plotted are each SNe Ibc subtype, along with the fit to the GRB-SN data, and the Pearson's correlation coefficient for that subtype.  No strong correlation is seen for any of the SN Ibc subtypes.}
\label{fig:SNe_Ibc}
\end{figure}

Next I considered the sample of non-GRB SNe Ibc originally presented in Cano (2013), where I determined $k$ and $s$ relative to SN~1998bw, to search for a similar correlation.  Only those events where the total extinction along the line of sight was known, and for which the host galaxy contribution was also removed, were included in the analysis.  Note that the stretch factors of the SNe Ibc are not corrected for redshift effects, however the bulk of the SNe Ibc considered here have redshifts of order $z\sim0.01$.  The resultant effect on the transformation between observer-frame to rest-frame for these very nearby events is of order a few percent, and thus does not fundamentally affect the results.  Plotted in Figure \ref{fig:SNe_Ibc} are the $k$- and $s$-factors of different SNe Ibc sub-types (Ib, Ic and Ic-BL), along with the best-fitting line, and 2-$\sigma$ uncertainty region found for the GRB-SNe sample.  Also displayed is the Pearson's correlation coefficient for each sample, as well as the combined SNe Ibc sample.  The low value of $r$ for each SN subtype highlights their unsuitability as standardizable candles.

\section{Discussion}

It has been noted that likely all core-collapse SNe possess a degree of asphericity (Wang \& Wheeler 2008), where the asphericity arises from, and is driven by, Rayleigh-Taylor instabilities surrounding the collapsing core.  This is in contrast to GRB-SNe, where asphericity also arises from the relativistic jet that is launched by the central engine.  Observationally, asphericity in SNe Ibc has been determined via different methods, such as modelling of optically thin nebular emission lines (e.g. Mazzali et al. 2005; Maeda et al. 2008; Taubenberger et al. 2009), where the emission lines of elements such as oxygen trace the overall velocity distribution of the ejecta. Another method is comparing the results of hydrodynamical explosion \& radiative transfer models with observations, in order to determine the ejecta geometry.  For example, it was shown that a large excess of luminosity along the polar/rotation axis relative to the equatorial axis (ratio of 4:1) was needed to explain the early bolometric LC of SN~1998bw (Maeda et al. 2006).  It is very probable that all GRB-SNe are highly aspherical (Woosley \& Bloom 2006; Maeda et al. 2006).  However, because we detect the initial $\gamma$-ray emission, as observers we must be near to the jet-axis, or the high-energy emission would not be observed.  The subsequent SN therefore must also be viewed close to the jet axis, meaning that we are observing each event at approximately the same viewing angle.  Conversely, because the non-GRB SNe Ibc are viewed at all possible viewing angles, and with each possessing a different degree of asphericity, it is not unexpected that no relation exists between the luminosity and stretch factors of these SNe Ibc relative to the template SN.  In the case of SN~2009bb, which was observed to be an engine-driven SN with a small amount of material moving at relativistic velocities similar to those of the low-energy GRB-SNe, the results here reinforce the supposition made by Pignata et al. (2011), that SN~2009bb was likely observed close to the polar/elongated axis.  It is tempting then to conclude that the possible use of GRB-SNe as standardizable candles arises, at least in part, from them possessing a similar ejecta geometry that is observed at roughly the same viewing angle.

Moreover, as the LCs of type I SNe (including GRB-SNe) are powered solely by the radioactive decay of $^{56}_{28}$Ni~$\rightarrow$~$^{56}_{27}$Co~$\rightarrow$ $^{56}_{26}$Fe (Arnett 1982; Woosley \& Weaver 1986), the fact that GRB-SNe have a distribution of brightnesses (i.e. $k$ values) results from each SN nucleosynthesizing a different amount of $^{56}$Ni.  If GRB-SNe are accurately described by the collapsar model, then the SN is expected to be driven by energy deposited by the jet as it disrupts the star, as well as by an energetic wind driven by viscous interactions in the accretion disk surrounding the newly formed compact object (MacFadyen et al. 1999).  Explosive nucleosynthesis of $^{56}$Ni occurs deep in the star as material displaced by the jet moves away at supersonic speeds, while further nucleosynthesis is driven by the wind from the accretion disk, with the latter possibly being the dominant producer of $^{56}$Ni for events such as SN~1998bw (MacFadyen et al. 2001).  A n\"aive conclusion to be drawn is that the correlation presented here suggests that a relation exists between the strength and energetics produced by the central engine and the resultant nucleosynthetic yields of $^{56}$Ni.  Moreover, the lack of correlation of $k$ and $s$ for the non-GRB SNe implies that the explosion and nucleosynthesis mechanism(s) are not correlated.

This n\"aive scenario does not account for the physical conditions of the SNe however.  In the context of SNe Ia, which are, of course, also standardizable candles, their LCs are also powered solely by the radioactive decay of nickel and cobalt, the amount of which determines the LC's peak brightness and width.  The width also depends on the photon diffusion time, which in turn depends on the physical distribution of the nickel in the ejecta, as well as the mean opacity of the ejecta.  In general, the opacity increases with increasing temperature and ionization (Woosley et al. 2007), thus implying that more nickel present in the ejecta leads to larger diffusion times.  This directly implies that fainter SNe Ia fade faster than brighter SNe Ia, thus satisfying the luminosity--decline relation (Phillips 1993).  This is not the only effect however, as the distribution of nickel in the ejecta also affects how the LC evolves, where nickel located further out has a faster bolometric LC decline.  Additionally, following maximum $B$-band light, SNe Ia colours are increasingly affected by the development of Fe II and Co II lines that blanket/suppress the blue $B$-band light.  Dimmer SNe are thus cooler, and the onset of Fe III $\rightarrow$ Fe II recombination occurs quicker than in brighter SNe Ia, resulting in a more rapid evolution to redder colours (Kasen \& Woosley 2007).  Therefore the faster $B$-band decline rate of dimmer SNe Ia reflects their faster ionization evolution, and provides additional clues as to why fainter SNe Ia fade more rapidly.  Thus, as the LCs of GRB-SNe are also powered by radioactive decay, the physics that govern SNe Ia also govern those of GRB-SNe, and may go some way to explaining why GRB-SNe are also standardizable candles.

\section{Conclusions}

In this paper I have shown that a statistically significant correlation exists between the brightness ($k$) and width/shape ($s$) of the LCs of GRB-SNe (the correlation coefficient being $r=0.932$) relative to a template SN (SN~1998bw).  This result highlights their potential use as standardizable candles.  Every single nearby, spectroscopically identified GRB-SN obeys this relation, including GRBs 980425, 030329, 031203, 060218, 100316D, 120422A and 130831A.  GRB~130427A has been omitted from this analysis due to its complicated LC structure, that will likely require more sophisticated decomposition such as that used by Deng et al. (2005) for GRB~030329.

When the relativistic, engine-driven type Ic-BL SN~2009bb (which is not associated with a GRB) was included in the analysis, it too was seen to follow this correlation ($r=0.916$ for the entire sample).  This highlights the possibility that perhaps all engine-driven SNe may eventually be used as standardizable candles, though this conjecture is very tentative as it is based on only a single event.

In this analysis, there are several sources of error that need to be considered.  These errors arise from the AG and host-galaxy subtractions, the uncertainties in the observer-frame and rest-frame extinction, and the SED interpolation.  In an attempt to properly account for the cumulative effect these will have on the resultant correlation between $k$ and $s$, I have adopted a conservative error of 20\% in the values of $k$ and $s$.  The adoption of this large error estimate is justified when compared with the original analysis of SNe Ia in Phillips (1993).  All nine SNe Ia in his sample were located at distances closer than SN~1998bw, which itself is the closest GRB-SN to date.  As such, the SNe Ia were all brighter than the GRB-SNe in the sample investigated here, and all had well-determined line-of-sight extinction values.  Moreover, no uncertainties were introduced from AG subtraction or SED construction and interpolation when determining the peak SN magnitudes and $\Delta m_{15}$ values.  This all implies that the luminosity--decline relation in Phillips (1993) is naturally better determined than the complicated decomposition technique employed here.

With one eye on the future, the modest size of the sample in this work highlights the need for additional observations of GRB-SNe, especially to obtain high-cadence and very importantly, multi-wavelength observations of each event.  The addition of further high-quality investigations of GRB-SNe will hopefully lead to a refinement and improvement of the scatter in the $k$--$s$ relationship, which in turn will allow for the possibility for their eventual use as a standardizable candle.  Once the redshift of a GRB has been determined, the resultant campaign to observe the accompanying SN should consider the rest-frame wavelengths and obtain observations to bracket as many rest-frame filters as possible.  Finally, but very importantly, in order to accurately determine the brightness and shape of the accompanying SN, it is necessary to have a good time coverage of the event at all epochs, as an incorrect AG model will give incorrect SN properties.

\acknowledgments

I am exceedingly grateful to Palli Jakobsson, Cristiano Guidorzi, David Alexander Kann and the anonymous referee for their comments and council on every aspect of this manuscript.  I happily acknowledge support by a Project Grant from the Icelandic Research Fund.

%{\it Facilities:} \facility{Nickel}, \facility{HST (STIS)}, \facility{CXO (ASIS)}.



\appendix

\section{Individual GRB-SNe}

Foreground extinction has been corrected for using the dust maps of Schlafly \& Finkbeiner (2011).

\subsection{GRB 011121 / SN 2001ke}

I used the host-subtracted optical data published in Bloom et al. (2002) and Garnavich et al. (2003).  A rest-frame extinction of $A_{V, \rm rest}=0.14\pm0.06$ mag, as determined by Kann et al. (2010), was used.  The data are well described by a single power-law (SPL), though the lack of early data limited the analysis.  After comparing with the values used in the literature, I used a decay rate of $\alpha=1.65\pm0.15$, which is similar to that found by Price et al. (2002; $\alpha=1.66$) and Garnavich et al. (2003; $\alpha=1.72$).  I also note that this value of $\alpha$ is similar to the pre-break value found by Greiner et al. (2003) of $\alpha_{1}=1.63\pm0.61$, who fit a BPL model to their data, finding a break time of $1.26\pm0.94$ d and a post-break decay of $\alpha_{2}=2.73\pm0.45$.

The derived values of $k$ are larger than those found by Ferrero et al. (2006), though the stretch factors agree well, which are all observer-frame $R$-band): $k_{R}=0.88\pm0.08$ and $s_{R}=0.80\pm0.02$.  And despite the differing late-time decay rate of the AG model used here and in Greiner et al. (2003), the latter find an observer-frame $R_{c}$-band luminosity factor of SN~2001ke of $k=0.85\pm0.11$.  

\subsection{GRB 030329 / SN 2003dh}

The optical LC of GRB 030329 has a very complex nature, with several authors attempting to decompose the thousands of observations obtained (e.g. Matheson et al. 2003; Lipkin et al. 2004; Deng et al. 2005).  I used the K-corrected $UBV$ LCs that resulted from the detailed decomposition of the optical spectral obtained by Matheson et al. (2003) that was performed by Deng et al. (2005), which were host-corrected.  Data in their Table 1 (for Case I) were used in this analysis.  Deng et al. (2005) corrected the spectra for foreground extinction only using the dust maps of Schlegel et al. (1998), which has the value $E(B-V)=0.025$ mag, while that of Schlafly \& Finkbeiner (2011) is $E(B-V)=0.021$ mag, implying a difference in extinction of $A_{V} \approx 0.011$ mag.  As the magnitudes determined by the decomposition by Deng et al. (2005) have errors ranging from 0.20--0.40 mag, this small difference is considered to be negligible.  Finally, I used a rest-frame extinction of $A_{V, \rm rest}=0.39\pm0.15$ mag as determined by Kann et al. (2006).

The derived values of $k$ and $s$ are consistent with those of Ferrero et al. (2006) within our respective errorbars.  Ferrero et al. (2006) find $k_{R}=1.50\pm0.19$ and $s_{R}=0.85\pm0.10$, and we both find that the luminosity factor is greater than the stretch factor.  

\subsection{GRB 031203 / SN 2003lw}

I used the photometry of the AG, SN and host published in Malesani et al. (2004).  The largest source of uncertainty is the unknown total extinction along the line of sight, where it is seen that GRB 031203 was heavily extinguished.  I adopted the result from the analysis of Margutti et al. (2007), who find a total extinction along the line of sight of $E(B-V)\approx1.1$ mag.  The dust maps of Schlafly \& Finkbeiner (2011) imply a foreground extinction of $E(B-V)_{\rm fore}\approx 0.90$ mag, leaving a rest-frame contribution of $E(B-V)_{\rm rest}\approx0.2$ mag, or $A_{V}\approx 0.65$ mag.  This highlights that most of the line-of-sight extinction is due to the MW.

The luminosity factors found here are somewhat brighter than those determined by Ferrero et al. (2006), who, for example, found $k_{R}=1.28\pm0.18$, though our stretch factors agree within their respective errorbars $s_{R}=1.09\pm0.07$.  Note that the analysis of Margutti et al. (2007) was performed after Ferrero et al. (2006) published their analysis, thus preventing a consistent analysis between the two datasets as different values for the entire line-of-sight extinction were used in each analysis.  My $k$ factors are also larger than published in Cano (2013), where I used the same the total line-of-sight extinction towards GRB 031203 as Ferrero et al. (2006).

\subsection{GRB 060218 / SN 2006aj}

I used the host-subtracted photometry published in Sollerman et al. (2006) and Ferrero et al. (2006), and $A_{V, \rm rest}\approx 0.13$ mag (Ferrero et al. 2006).  It was noted by Ferrero et al. (2006) that an additional power-law component was needed to fit their LCs, where they found that the decay rate ($\alpha$) decreased in successively redder filters (and thus contributed more in the redder filters).  This behaviour can also be seen in the \emph{Swift} observations of GRB 060218 (Campana et al. 2006), where the rate of decay of the suspected shock-heated material is greater at bluer wavelengths.  As such I have also included a SPL component in our fit, where I also find that the value of alpha increases in redder filters.  It is worth noting that Ferrero et al. (2006) fit the LC of SN~2006aj both with and without a SPL component, and in both cases the stretch factor varied by a small amount ($\Delta s\sim 0.02$--0.04), while the luminosity factor increased by no more than $\Delta k=0.1$.  Thus in both cases (including a SPL or not) the values of $k$ and $s$ follow the correlation found in this paper.

The derived values of $k$ and $s$ agree excellently with the observer-frame values found by Ferrero et al. (2006) and Cano et al. (2011A).  It appears that both $s$ and $k$ vary in the range $\approx$0.6--0.75.  I note as well that the good agreement between the rest-frame and observer-frame values of $k$ and $s$ in each observer-frame/rest-frame filter is not unexpected due to the low redshift of GRB 060218 ($z=0.033$).

\subsection{GRB 090618}

I repeated the analysis I presented in Cano et al. (2011B), using the already published host-subtracted magnitudes.  GRB 090618 is the only event in the sample where a broken power-law (BPL) was needed to fit the early AG, where the decay constant before and after the break time are consistent with the analysis in Cano et al. (2011B), albeit with a slightly later break time.  Moreover, the values of $k$ and $s$ are in excellent agreement with the previous analysis.

\subsection{GRB 100316D / SN 2010bh}

I used the host-subtracted photometry presented in Olivares et al. (2012), as well as their determination of the rest-frame extinction, $A_{V, \rm rest}\approx1.2$ mag, which was determined via SED fitting of the early AG, from optical to X-ray wavelengths.  Note, however, that this rest-frame extinction is larger than that determined via the sodium doublet absorption features in the spectra of Bufano et al. (2012), which is similar to that determined in Cano et al. (2011A), who followed the SN LC analysis derived by Drout et al. (2011).  It is clear that the value of the rest-frame extinction is an on-going a matter of debate and uncertainty.

When fitting the data, I included an additional PL component.  Both Olivares et al. (2012) and Cano et al. (2011A) noted that the early emission from GRB~100316D was brighter than GRB~980425 at a similar epoch, prompting me to suggest in Cano et al. (2011A) that the early $B$-band emission originated from shock-heated material, similar to that seen for GRB~060218.  When the larger rest-frame extinction and the SPL was included in the analysis, the values of $k$ and $s$ were seen to be in very good agreement with those determined in Cano et al. (2011A).  Note that the analysis presented in Cano et al. (2011A) to find $k$ and $s$ did not include a SPL component.

Recent work by Lyman et al. (2014) has found that the peak bolometric magnitude, and the $\Delta m_{15}$ value of GRB 100316D are very similar to that of GRB 060218, implying that the stretch and luminosity factors of the two events are approximately similar, even when accounting for different colour evolutions.  Further evidence for the similar shape and brightness of GRBs 060218 and 100316D arise from the fact that independent analyses find that similar amounts of $^{56}$Ni were nucleosynthesized: $0.1\ M_{\odot} \le M_{\rm Ni} \le 0.2\ M_{\odot}$ (Cano et al. 2011A; Olivares et al. 2012; Bufano et al. 2012). Lyman et al. (2014) find a larger $^{56}$Ni for 2010bh of $\approx0.35\ M_{\odot}$, however they also find that SN~2006aj nucleosynthesized the same amount of $^{56}$Ni, further supporting the similarities in their LCs.  Admittedly, the values of $k$ and $s$ are less certain in this event, though there is support that their values are similar to SN~2006aj.

\subsection{GRB 120422A / SN 2012bz}

I used the photometry of the AG, SN and host published in Schulze et al. (2014).  Both Schulze et al. (2014) and Melandri et al. (2012) found a negligible rest-frame extinction (i.e. $A_{V, \rm rest} \approx 0$ mag), which we also adopt.  I also included a SPL component in the analysis, which was also included in the analysis performed by Schulze et al. (2014).  I note that the values of $k$ and $s$ are in excellent agreement with those found by Schulze et al. (2014).  For example, at $z=0.283$ observer-frame $r$ is roughly rest-frame $V$, where I find $k_{V, \rm rest}=1.27\pm0.03$ and $s_{V, \rm rest}=0.91\pm0.02$, while Schulze et al. (2014) find  $k_{r, \rm obs}=1.25\pm0.02$ and $s_{r, \rm obs}=0.90\pm0.01$.

\subsection{GRB 130831A / SN 2013fu}

I used the photometry of the AG, SN and host recently published in Cano et al. (2014), using the value of the rest-frame extinction of, $A_{V, \rm rest}=0.21^{+0.28}_{-0.21}$ mag.  A SPL fits the data well, and not unsurprisingly I find an identical decay rate, $k$ and $s$ values as in Cano et al. (2014).

\clearpage

\clearpage

\begin{table}
\begin{center}
 \caption{GRB-SNe -- vital statistics}
\label{table:refs}
\medskip
\begin{tabular}{ccccccc}
\hline
GRB & SN & $z$ & A$_{V,\rm fore}$ (mag) & A$_{V,\rm rest}$ (mag) & Filters (obs) & Ref.\\
\hline
980425	&	1998bw	&	0.0085	&	0.16	$\pm$	0.00	&	$^{\dagger}$			&	$UBVRI$	&	(1)	\\
011121	&	2001ke	&	0.36	&	1.31	$\pm$	0.04	&	0.39	$\pm$	0.14	&	$VRI$	&	(2,3)	\\
030329	&	2003dh	&	0.1685	&	0.06	$\pm$	0.00	&	0.39	$\pm$	0.15	&	$UBV$	&	(4)	\\
031203	&	2003lw	&	0.1055	&	2.82	$\pm$	0.10	&	$\approx 0.65$			&	$VRIJ$	&	(5,6)	\\
060218	&	2006aj	&	0.033	&	0.39	$\pm$	0.01	&	0.13	$\pm$	0.00	&	$UBVRI$	&	(7,8)	\\
090618	&	-	&	0.54	&	0.23	$\pm$	0.01	&	0.29	$\pm$	0.00	&	$Ri$	&	(9)	\\
100316D	&	2010bh	&	0.059	&	0.31	$\pm$	0.00	&	$\approx 1.2$			&	$grizJ$	&	(10)	\\
120422A	&	2012bz	&	0.283	&	0.09	$\pm$	0.00	&	0.00	$\pm$	0.00	&	$gri$	&	(11)	\\
130831A	&	2013fu	&	0.479	&	0.12	$\pm$	0.00	&	0.21	$\pm$	0.21	&	$Ri$	&	(12)	\\
\hline
\end{tabular}
\begin{flushleft}
$^{\dagger}$ Rest-frame extinction was determined on a time-dependent basis by \cite{Clocchiatti2011}.\\
\textbf{References}: (1) \cite{Clocchiatti2011}; (2)\cite{Bloom2002}; (3) \cite{Garnavich2003}; (4)~\cite{Deng2005}; (5) \cite{Malesani2004}; (6) \cite{Margutti2007}; (7) \cite{Sollerman2006}; (8) \cite{Ferrero2006}; (9) \cite{Cano2011_090618}; (10) \cite{Olivares2012}; (11)~\cite{Schulze2014}; (12) \cite{Cano2014trio}.
\end{flushleft}
\end{center}
\end{table}

\begin{table}
\small
\centering
\caption{GRB-SNe -- Best-fitting parameters}
\label{table:params}
\medskip
\begin{tabular}{cccccccc}
\hline
GRB	&	Filter (rest)	&	model$^{a}$	&	$\alpha{1}$	&	$\alpha{2}$	&	$T_{\rm B}$ (d)	&	$k$	&	$s$	\\
\hline
011121	&	$B$	&	SPL	&	$1.65\pm0.15$	&	-	&	-	&	$1.21\pm0.24$	&	$0.82\pm0.16$	\\
	&	$V$	&	SPL	&	$1.65\pm0.15$	&	-	&	-	&	$1.04\pm0.21$	&	$0.85\pm0.17$	\\
030329	&	$U$	&	SN	&	-	&	-	&	-	&	$1.21\pm0.24$	&	$0.90\pm0.18$	\\
	&	$B$	&	SN	&	-	&	-	&	-	&	$1.25\pm0.25$	&	$0.86\pm0.17$	\\
	&	$V$	&	SN	&	-	&	-	&	-	&	$1.39\pm0.28$	&	$0.86\pm0.17$	\\
031203	&	$V$	&	SN	&	-	&	-	&	-	&	$1.82\pm0.36$	&	$1.19\pm0.24$	\\
	&	$R$	&	SN	&	-	&	-	&	-	&	$1.69\pm0.34$	&	$1.15\pm0.23$	\\
	&	$I$	&	SN	&	-	&	-	&	-	&	$1.45\pm0.29$	&	$0.96\pm0.19$	\\
060218	&	$U$	&	SPL	&	$1.11\pm0.08$	&	-	&	-	&	$0.54\pm0.11$	&	$0.63\pm0.13$	\\
	&	$B$	&	SPL	&	$0.86\pm0.25$	&	-	&	-	&	$0.52\pm0.10$	&	$0.63\pm0.13$	\\
	&	$V$	&	SPL	&	$0.33\pm0.31$	&	-	&	-	&	$0.60\pm0.12$	&	$0.69\pm0.14$	\\
	&	$R$	&	SPL	&	$0.49\pm0.70$	&	-	&	-	&	$0.65\pm0.13$	&	$0.72\pm0.14$	\\
090618	&	$B$	&	BPL	&	$0.72\pm0.19$	&	$2.04\pm0.12$	&	$1.26\pm0.49$	&	$1.11\pm0.22$	&	$0.98\pm0.20$	\\
100316D	&	$V$	&	SPL	&	$0.03\pm0.01$	&	-	&	-	&	$0.56\pm0.15$	&	$0.53\pm0.11$	\\
	&	$R$	&	SPL	&	$0.01\pm0.02$	&	-	&	-	&	$0.56\pm0.11$	&	$0.56\pm0.11$	\\
	&	$I$	&	SPL	&	$0.02\pm0.02$	&	-	&	-	&	$0.48\pm0.10$	&	$0.49\pm0.10$	\\
120422A	&	$B$	&	SPL	&	$0.70\pm0.07$	&	-	&	-	&	$1.00\pm0.20$	&	$0.92\pm0.18$	\\
	&	$V$	&	SPL	&	$0.78\pm0.11$	&	-	&	-	&	$1.26\pm0.25$	&	$0.93\pm0.19$	\\
130831A	&	$B$	&	SPL	&	$1.64\pm0.10$	&	-	&	-	&	$0.95\pm0.19$	&	$0.82\pm0.19$	\\
\hline
\end{tabular}
\begin{flushleft}
$^{a}$ SPL$=$Single power-law; BPL$=$Broken power-law; SN$=$No AG, SN only.\\

\end{flushleft}
\end{table}

\end{document}